\documentclass[prb,twocolumn,floatfix,showpacs,superscriptaddress]{revtex4}
\usepackage{graphics}
\usepackage{epsfig}
\usepackage{times}
\usepackage{bm}
\usepackage{color}
\usepackage{amsmath}	
\begin{document}
\title{Tomonaga-Luttinger-liquid criticality: numerical entanglement entropy approach}
\author{Satoshi Nishimoto}
\affiliation{Institut f\"ur Theoretische Festk\"orperphysik, IFW Dresden, D-01171 Dresden, Germany}

\date{\today}
\begin{abstract}
The von Neumann entanglement entropy is studied with the density-matrix renormalization 
group technique. We propose a simple approach to calculate the central charge using 
the entanglement entropy for one-dimensional (1D) quantum system. This approach is applied 
to a couple of quantum systems: (i) 1D frustrated spin model and (ii) 1D half-filled spinless fermions 
with nearest-neighbor repulsion; and, it is confirmed that the central charge is estimated 
very accurately for the both systems. Also, a new method to determine the critical point between 
TL-liquid and gapped (or ordered) phases from the proposed approach is suggested. 
Furthermore, we mention that the Tomonaga-Luttinger parameter can be obtained in a like 
manner as the central charge, using the charge-density fluctuation of a part of the 1D system.
\end{abstract}
\maketitle

\section{Introduction}

Quantum entanglement~\cite{Horodecki09} is a quantum mechanical phenomenon 
created by the separation of interacting quantum particles.  This fundamental concept underpins 
a wide range of research in physics, such as quantum information, quantum computing, 
and quantum gravity, etc. When one considers this phenomenon in a study, it is often 
quantified as an {\it entanglement entropy}. In the field of condensed-matter physics, 
the von Neumann entanglement entropy has been frequently used as an order parameter 
to investigate quantum phase transition or topological order in the quantum many-body problems. 
In particular, for one-dimensional (1D) quantum systems the entropy can be directly related 
to the {\it central charge} of the conformal field theory (CFT).~\cite{Affleck91,Holzhey94} 
To date, a method with this relation has been developed as a powerful tool to study 
the universal properties of the Tomonaga-Luttinger (TL) liquid~\cite{Giamarchi04}, in combined with numerical 
methods such as exact diagonalization (ED) or density-matrix renormalization group (DMRG)~\cite{White92}  
techniques. 

Let us consider a quantum 1D periodic system with length $L$. 
The von Neumann entanglement entropy of its subsystem with length $l$ is given as
$S_L(l)=-{\rm Tr}_l \rho_l \log \rho_l$, where $\rho_l={\rm Tr}_{L-l}\rho$ is 
the reduced density matrix of the subsystem and $\rho$ is the full density matrix of 
the whole system. Using the CFT, the entropy of the subsystem with length $l$ 
for a fixed system length $L$ has been derived:~\cite{Affleck91,Holzhey94,Calabrese04}
\begin{eqnarray}
S_L(l)=\frac{c}{3}\ln\left[\frac{L}{\pi}\sin\left(\frac{\pi l}{L}\right)\right]+s_1
\label{entropy}
\end{eqnarray}
where $c$ is the central charge of the associated CFT and $s_1$ is a non-universal constant. 
A prime objective of using this formula is to estimate the central charge,~\cite{Laflorencie06,Legeza07} which provides 
definitive information concerning the universality class of $(1+1)$ dimensional system.~\cite{Cardy96} 
This estimation has been applied to a variety of fermionic systems; e.g., with impurity~\cite{Zhao06, Ren09,Refael09}, 
coupled to bosons,~\cite{Lauchli08,Ejima11} and under the magnetic field,~\cite{Manmana11} etc. 
Although the central charge can be also obtained by examining the finite-size correction 
of ground-state energy or the low-temperature behavior of specific heat,~\cite{Bloete86} 
excited states have to be taken into account and considerable difficulty is involved in the ED 
and DMRG calculations. But, if one uses Eq.(\ref{entropy}), it is only necessary to consider 
the ground state. 

So far, two kinds of numerical methods have been principally adopted to extract 
the value of $c$ from Eq.(\ref{entropy}): one is direct fitting of $S_L(l)$ as a function of $l$ 
for a fixed system length and the other is scaling analysis using a relation 
$\Delta S \equiv S_L(\frac{L}{2})-S_{L^\prime}(\frac{L^\prime}{2}) \simeq \frac{c}{3}\ln \frac{L}{L^\prime}$, 
considering two systems with lengths $L$ and $L^\prime$. However, one would encounter 
difficulty attributable to a finite-size effect in their practical use. More specifically, 
in the former method the equation (\ref{entropy}) is exact only in the thermodynamic limit 
$L \to \infty$ and some ambiguity remains in the fitting of finite-size result;~\cite{Franca08} 
whereas, in the latter method the system length to be studied must be strongly restricted 
due to the incommensurability (or frustration) of spin and/or charge fluctuations  as well as 
the (open-)shell problem in itinerant systems. Therefore, the aim of this paper is to propose 
a new method that overcomes those problems. We first derive an efficient formula for calculating 
the central charge. And then, to ascertain the validity of the formula we apply it to a couple 
of 1D quantum systems; (i) frustrated spin model and (ii) itinerant model with strong charge 
fluctuation. We demonstrate that a very accurate estimation of critical point between 
TL-liquid and gapped (or ordered) states is enabled with this method. Furthermore, 
we suggest that the TL parameter can be calculated in a like manner as the central charge.

\section{Derivation of the formula}

Let us now derive the formula for estimating the central charge. We first prepare $S_L(\frac{L}{2})$ 
and $S(\frac{L}{2}-1)$ from Eq.(\ref{entropy}). Then, by subtracting the one from the other 
a simple expression is obtained: 
\begin{eqnarray}
c=\frac{3\left[S_L\left(\frac{L}{2}-1\right)-S_L\left(\frac{L}{2}\right)\right]}
{\ln\left[\cos\left(\frac{\pi}{L}\right)\right]} \equiv c_1.
\label{cc1}
\end{eqnarray}
Using this expression we can easily calculate the central charge for a 1D system with a fixed system length. 
The main advantages of Eq.(\ref{cc1}) are; (i) the non-universal constant $s_1$ does not appear explicitly, 
(ii) only two values of $S_L(l)$ at the middle of system, where the finite-size correction to 
Eq.(\ref{entropy}) is the smallest~\cite{Franca08} and the DMRG calculation is the most accurate, 
are used, and (iii) a unique value of $c$ is estimated for each system length and it enables us to 
perform a systematic extrapolation to the thermodynamic limit. Also, an alternative 
expression can be derived from the second derivative of Eq.(\ref{entropy}) with respect to $l$. 
Since $S_L(l)$ is symmetric about $l=\frac{L}{2}$, i.e., $S_L(l)=S_L(L-l)$, we obtain
\begin{eqnarray}
\nonumber
c&=&-\frac{3L^2}{\pi^2}\frac{\partial^2S_L(l)}{\partial l^2}\bigg|_{l=\frac{L}{2}}\\
&\simeq&\frac{6L^2}{\pi^2}\left[S_L\left(\frac{L}{2}\right)-S_L\left(\frac{L}{2}-1\right)\right] \equiv c_2.
\label{cc2}
\end{eqnarray}
Note that Eq.(\ref{cc2}) can be also derived by applying 
$\ln\left[\cos\left(\frac{\pi}{L}\right)\right]\simeq-\frac{\pi^2}{2L^2}$ into Eq.(\ref{cc1}).

\begin{table}[tb]
\caption{System-size dependence of $c_1$ and $c_2$ obtained from Eqs.(\ref{cc1}) and 
(\ref{cc2}) for the 1D $S=\frac{1}{2}$ frustrated Heisenberg model.}
\begin{center}
\begin{tabular}{cccccc}
\hline
\hline
            &  \multicolumn{2}{c}{$J_2/J_1=0$} &  \multicolumn{2}{c}{$J_2/J_1=0.2$} \\
     $L$ &  $c_1$  & $c_2$ &  $c_1$  & $c_2$ & \\
\hline
     32  &   1.005297   &   1.006916   &   1.001593   &   1.003206   &  \\
     64  &   1.002932   &   1.003335   &   1.000454   &   1.000856   &  \\
     96  &   1.002207   &   1.002386   &   1.000241   &   1.000420   &  \\
    128 &   1.001845   &   1.001945   &   1.000164   &   1.000260   &  \\
\hline
\hline
 \end{tabular}
\end{center}
  \label{ccJJ}
\end{table}

\section{Application to quantum systems}

\subsection{frustrated spin chain}

In order to confirm the validity of Eqs.(\ref{cc1}) and (\ref{cc2}), we apply them to a couple 
of 1D quantum systems. The first system considered is the $S=\frac{1}{2}$ frustrated Heisenberg 
chain. The Hamiltonian is given by
\begin{eqnarray}
H = J_1 \sum_i \vec{S}_i \cdot \vec{S}_{i+1} + J_2 \sum_i \vec{S}_i \cdot \vec{S}_{i+2},
\label{hamJJ}
\end{eqnarray}
where $\vec{S}_i$ is a spin-$\frac{1}{2}$ operator at site $i$. The parameters $J_1$ and $J_2$ 
are nearest-neighbor and next-nearest-neighbor antiferromagnetic exchange interactions, 
respectively.  This system has been extensively studied both analytically and 
numerically:~\cite{Majumdar69,Haldane82,Tonegawa87,Okamoto92,Eggert96,White96} 
the ground state is of a dimerized zigzag-bond state for $0.241 \lesssim J_2 /J_1 \le 0.5$ 
and of the Majumdar-Ghosh state with incommensurate spiral correlations for 
$J_2/J_1 \ge 0.5$; accordingly, the spin gap opens when $J_2/J_1 \gtrsim 0.241$. 
While for $J_2/J_1 \lesssim 0.241$, the system is described as a TL liquid and the central charge 
is expected to be unity ($c=1$). In recent years this system has been frequently used as a touchstone 
of numerical methods.~\cite{Alet10,Thomale10,Dalmonte11,Rachel11}

By the DMRG method we study the systems with lengths $L=32$ to $144$ keeping $m \sim 25L$ 
density-matrix eigenstates. Note that an appropriate 1D array for the construction of the PBC 
is necessary to obtain highly-accurate result (see Fig. 1 of Ref.~\onlinecite{Qin95}). 
In this way, the largest discarded weight is $w_{\rm d} \sim 2 \times 10^{-12}$ 
in the renormalization procedure and the central charge converges at least six digits as a function 
of $m$ for a fixed system length. As examples, the DMRG results of $c_1$ and $c_2$ at $J_2/J_1 =0$ 
and $0.2$, where the system is  in the TL-liquid phase, are shown in Table~\ref{ccJJ} for several 
system lengths. We can see that both of the quantities  $c_1$ and $c_2$ converge very fast with 
increasing the system length and they can be easily extrapolated to $1$ in the thermodynamic limit. 
In fact, even for $L=32$ the largest deviations from $c=1$ are only $0.5 \%$ and $0.7 \%$ for 
$c_1$ and $c_2$, respectively. A finite-size correction is always smaller in $c_1$ than in $c_2$, 
so that we make use only of $c_1$, namely Eq.(\ref{cc1}), for the finite-size-scaling analysis below.

\begin{figure}[t]
\centering
\includegraphics[clip,scale=0.5]{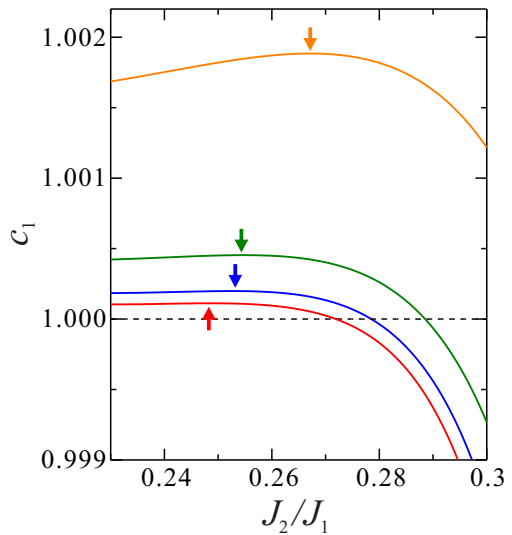}
\caption{(Color online) System-size dependence of $c_1$ as a function of $J_2/J_1$. The arrows indicate 
maximum positions $(J_{\rm 2,max}/J_1, c_{\rm 1,max})$. The system length is $L=32$ (red), $64$ (blue), 
$96$ (green), and $128$ (orange) from top to bottom. 
}
\label{fig1}
\end{figure}

\begin{figure}[t]
\centering
\includegraphics[clip,scale=0.5]{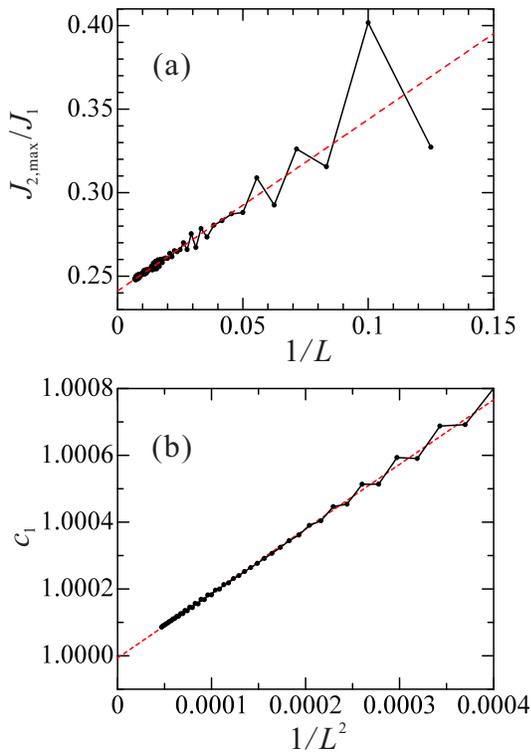}
\caption{(Color online) (a) Maximum position $J_{\rm 2,max}/J_1$ as a function of $1/L$ for systems with 
even number of sites from $L=32$ to $L=144$. The red dashed line denotes a linear least-square fit 
with a weight function $w(1/L)=L$. (b) Maximum height $c_{\rm 1,max}$ as a function of $1/L^2$ 
for systems with even number of sites from $L=32$ to $L=144$. The red dashed line shows a simple 
linear fitting.
}
\label{fig2}
\end{figure}

Now we demonstrate how one can estimate a critical point between the TL- and non-TL-liquid 
(or gapped) phases with Eq.(\ref{cc1}). This is a good enduring test of Eq.(\ref{cc1}) because 
the use of $\Delta S$ is no longer allowed due to the system-size dependent frustration 
of the system (\ref{hamJJ}). This system belongs to the Gaussian universality class ($c=1$) 
for the TL-liquid phase and $c<1$ is expected for the gapped phase from the renormalization 
in the massive region.~\cite{Inoue99} In Fig.~\ref{fig1} the DMRG results of $c_1$ is plotted 
as a function of $J_2/J_1$ for several system lengths. In the renormalization group the correction 
of central charge is expressed as $c=1+{\cal O}(f^3)$ where $f$ is the coupling constant 
of marginal operators,~\cite{Fateev93,Calabrese10} so that a `transition point' for a fixed system length 
may be given by a maximum of $c_1$. Actually, we find that all of those curves have a maximum at 
some point $(J_{\rm 2,max}/J_1, c_{\rm 1,max})$; on the left-hand side of the maximum point 
they becomes flatter and flatter, and gets closer to $c_1=1$ with increasing the system length; 
whereas on the right-hand side, $c_1$ drops down abruptly as a function of $J_2/J_1$. 
Hence, the maximum position must approach the critical point $J_2=J_{\rm 2,c}$ as the system 
length increases and reach there with $c_1=1$ in the thermodynamic limit; namely, 
$J_{\rm 2,max} \to J_{\rm 2,c}$ and $c_{\rm 1,max} \to 1$ as $L \to \infty$. 

In Fig.~\ref{fig2}(a) the maximum position $J_{\rm 2,max}/J_1$ is plotted as a function 
of $1/L$ for systems with even number of sites from $L=32$ to $144$. Although it oscillates 
due to the inconsistency between the periodicity of spin wave and the system length, i.e.,  
system-size dependent frustration, its amplitude decreases rapidly with increasing system length. 
Therefore, an extrapolation of $J_{\rm 2,max}/J_1$ to the thermodynamic limit is practicable and 
we obtain $J_{\rm 2,c}/J_1=0.24112778$ by linear least-square fit with a weight function 
$w(1/L)=L$ [without the weighting, $J_{\rm 2,c}/J_1=0.2414 \pm 0.001$]. This value is consistent 
with the previous estimations.~\cite{Okamoto92,Eggert96} Another thing to be examined is an extrapolation 
of  the maximum height to the thermodynamic limit. The maximum height $c_{\rm 1,max}$ is 
plotted as a function of $1/L^2$ in Fig.~\ref{fig2}(b). Interestingly, the data seem to be on 
a straight line at least for longer systems and we obtain $c=0.999992824$ in the thermodynamic 
limit by a linear fitting. It implies that the central charge scales as 
\begin{eqnarray}
c_1=1+{\cal O}(1/L^2).
\label{corrections}
\end{eqnarray}
This can be interpreted as the ${\cal O}(1/L^2)$ corrections originated from the $x=4$ irrelevant 
fields in the CFT.~\cite{Cardy86} No sooner $J_2$ moves away from $J_{\rm 2,c}$ than 
the system-size dependence of $c_1$ starts to deviate from Eq.(\ref{corrections}). 
It means that the logarithmic corrections are completely eliminated only at this critical point. 
It is consistent with the fact that the effective Hamiltonian of the system (\ref{hamJJ}) is purely 
Gaussian only at the critical point. In this manner we can easily determine whether or not the logarithmic 
corrections are present for arbitrary 1D quantum system.

\begin{table}[tb]
\caption{System-size dependence of $c_1$ and $c_2$ obtained from Eqs.(\ref{cc1}) and (\ref{cc2}) 
for the 1D half-filled spinless model with nearest-neighbor repulsion.}
\begin{center}
\begin{tabular}{cccccc}
\hline
\hline
          &  \multicolumn{2}{c}{$V/t=0$} &  \multicolumn{2}{c}{$V/t=1.9$} \\
     $L$ &  $c_1$  & $c_2$ &  $c_1$  & $c_2$ & \\
\hline
   30   &   1.001119   &   1.002954    &    1.005450   &   1.007293  &  \\
   70   &   1.000199   &   1.000535    &    1.002542   &   1.002879  &  \\
  110  &   1.000082   &   1.000218    &    1.001818   &   1.001955  &  \\
  150  &   1.000041   &   1.000114    &    1.001467   &   1.001540  &  \\
\hline
\hline
\end{tabular}
\end{center}
\label{cctV}
\end{table}

\begin{figure}[t]
\centering
\includegraphics[clip,scale=0.5]{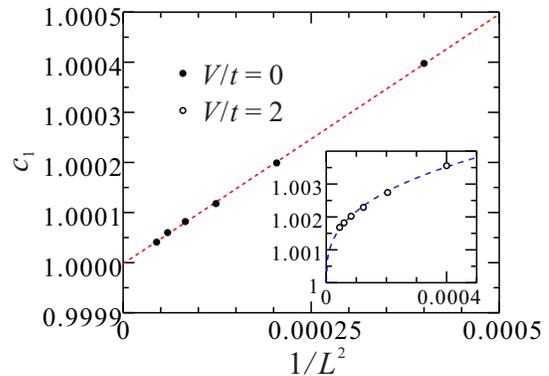}
\caption{(Color online) System-size dependence of $c_1$ as a function of $1/L^2$ for $V/t=0$. 
The system lengths are taken at intervals of $20$ sites from $L=50$ to $L=150$. 
The red dashed line shows a simple linear fitting. Inset: similar figure as the main one for $V/t=2$. 
The blue dashed line is guide to the eyes only.
}
\label{fig3}
\end{figure}

\begin{figure}[t]
\centering
\includegraphics[clip,scale=0.5]{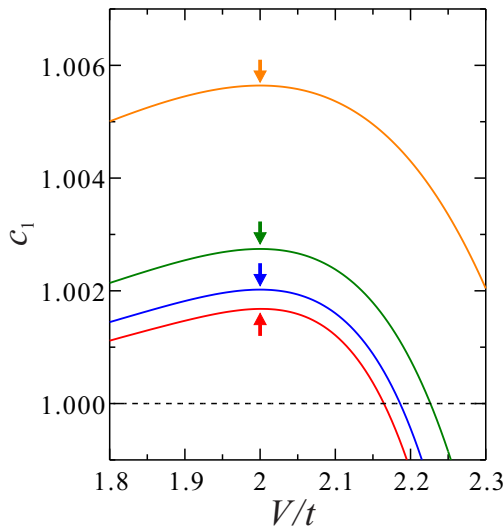}
\caption{(Color online) Finite-size-scaling analysis of $c_1$ as a function of $V/t$. The arrows indicate 
maximum positions $(V_{\rm max}/t, c_{\rm 1,max})$. The system length is $L=30$ (red), 
$70$ (blue), $110$ (green), and $150$ (orange) from top to bottom.
}
\label{fig4}
\end{figure}

\subsection{spinless fermions with nearest-neighbor repulsion}

It would be of importance to examine whether Eqs.(\ref{cc1}) and (\ref{cc2}) are applicable 
to itinerant fermion systems. Thus, as our second test we consider the 1D half-filled spinless fermions 
with nearest-neighbor repulsion. The Hamiltonian is given by
\begin{eqnarray}
H = -t \sum_{\langle i,j \rangle} (c_i^\dagger c_j + {\rm h.c.}) + V \sum_{\langle i,j \rangle} n_i n_j,
\label{hamtV}
\end{eqnarray}
where $c_i^\dagger$ ($c_i$) is a creation (annihilation) operator of a spinless fermion 
and $n_i$ ($=c_i^\dagger c_i$) is the corresponding number operator. The repulsive interaction 
$V$ ($>0$) is assumed to act only between neighboring sites $\langle i,j \rangle$. The particle 
density is fixed at $n=1/2$. This system can be mapped onto the exactly solvable Heisenberg XXZ chain 
and it is known that a transition between TL-liquid metallic and charge-density-wave insulating phases 
occurs at $V/t = 2 (\equiv  V_{\rm c}/t)$. 

We study the systems with lengths $L=30$ to $150$ keeping $m \sim 40L$ density-matrix eigenstates. 
This way, the largest discarded weight is $w_{\rm d} \sim 8 \times 10^{-12}$ in the renormalization procedure and 
the central charge converges at least seven digits as a function of $m$ for a fixed system length. 
Table~\ref{cctV} shows the DMRG results of $c_1$ and $c_2$ at $V/t =0$ and $1.9$, where 
the system (\ref{hamtV}) is in the TL-liquid phase, for several system lengths. 
In common with the case of the frustrated spin chain, the convergence of $c_1$ and $c_2$ with 
the system length is very fast and the deviation from $c=1$ is always below $1 \%$ for $L>30$. 
In Fig.~\ref{fig3} we plot the values of $c_1$ as a function of $1/L^2$ for some values of $V/t$. 
At $V/t=0$, we see that $c_1$ is scaled by Eq.(\ref{corrections}) and extrapolated to $c_1=1.00000062$ 
in the thermodynamic limit; however, for $V/t>0$ Eq.(\ref{corrections}) is no longer fulfilled 
due to the occurrence of the logarithmic corrections. As an example, the finite-size scaling of 
$c_1$ at $V/t=2$, where the logarithmic corrections are maximum, is shown in the inset of Fig.~\ref{fig3}. 

So, let us check the behavior of $c_1$ on $V/t$ near the critical point. In Fig.~\ref{fig4} the values 
of $c_1$ is plotted as a function of $V/t$ for several system lengths. As seen in the frustrated spin chain, 
each of the curves have a maximum at a point $(V_{\rm max}, c_{\rm 1,max})$. 
Note that the peak position is exactly $V_{\rm max}/t=2$ for all system lengths. This is because 
the `critical point' is independent of the system length in this model (\ref{hamtV}). Quite interestingly, 
it means that the `critical point' is not affected by the logarithmic corrections. Thus, we decide that Eqs.(\ref{cc1}) 
and (\ref{cc2}) are useful for 1D itinerant system as well. Separately from this paper, it has been 
verified that this method is successfully applied to a 1D spinless fermions with boson affected hopping.~\cite{Fehske11}

\begin{table}[t]
\caption{System-size dependence of $K_\rho(L)$ obtained from Eqs.(\ref{KrhoL}) 
for the 1D half-filled spinless model with nearest-neighbor repulsion. The extrapolated 
values to the thermodynamic limit ($L \to \infty$) and the exact values [Eq.(\ref{Krhoexact})] 
are also shown.
}
\begin{center}
\begin{tabular}{cccccc}
\hline
\hline
     $L$ &  $V=0$  & $V=1.8$ &  $V=1.9$  & \\
\hline
   30                 &   0.9926566   &  0.5212309   &  0.4940692   &  \\
   50                 &   0.9973670   &  0.5476091   &  0.5202473   &  \\
   70                 &   0.9986855   &  0.5587141   &  0.5314329   &  \\
   90                 &   0.9992341   &  0.5648065   &  0.5376011   &  \\
   110               &   0.9995373   &  0.5683950   &  0.5417063   &  \\
$L \to \infty$  &   1.0000593   &   0.5843504   &  0.5596052   &  \\
Exact               &   1.0000000   &   0.5838163   &  0.5562247   &  \\
\hline
\hline
\end{tabular}
\end{center}
\label{Krho}
\end{table}

\begin{figure}[t]
\centering
\includegraphics[clip,scale=0.5]{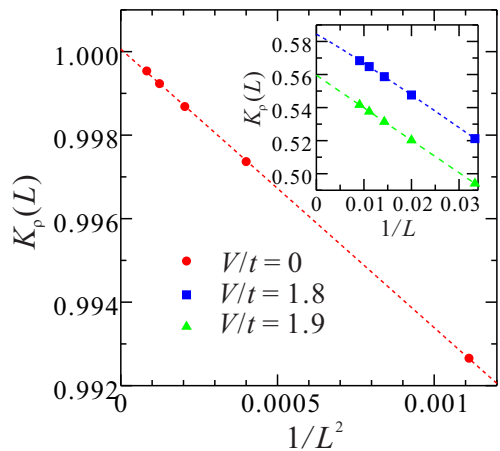}
\caption{(Color online) Finite-size-scaling analysis of $K_\rho(L)$ as a function of $1/L^2$ for $V/t=0$. 
The system lengths are taken at intervals of $20$ sites from $L=30$ to $L=110$. 
The red dashed line shows a simple linear fitting. Inset: finite-size-scaling analysis of 
$K_\rho(L)$ as a function of $1/L$ for $V/t=1.8$ and $1.9$. The blue and green dashed 
lines are fitting with function $K_\rho(L)=K_\rho+aL^{-\gamma}$. 
Through all the calculations $L/2$ is kept to be odd, and therefore, $a<0$ (see text).
}
\label{fig5}
\end{figure}

Lastly, we mention that the TL parameter $K_\rho$, which is one of the most notable quantity to 
study the TL-liquid properties, can be estimated in a manner similar to above. The TL parameter is 
related to the charge-density fluctuation of subsystem with length $l$, 
$F_L(l)=\langle(\sum_i n_i-\sum_i \bar{n}_i)^2\rangle$  [$i \in$ subsystem],
like~\cite{Song10}
\begin{eqnarray}
\pi^2F_L(l)=K_\rho\ln\left[\frac{L}{\pi}\sin\left(\frac{\pi l}{L}\right)\right]-\frac{(-1)^lA}{\left[\frac{L}{\pi}\sin\left(\frac{\pi l}{L}\right)\right]^{2K_\rho}}+f_1,
\label{fluctuation}
\end{eqnarray}
where $A$ and $f_1$ are non-universal constants. The function $F_L(l)$ behaves similarly to $S_L(l)$ with $l$. 
Then, in the same way as Eq.(\ref{cc1}) is derived, we easily obtain for $L \gg 1$
\begin{eqnarray}
\nonumber
&&K_\rho\left[1+(-1)^\frac{L}{2}2A\left(\frac{\pi}{L}\right)^{2K_\rho}\right] \\
&&=\frac{\pi^2\left[F_L\left(\frac{L}{2}-2\right)-F_L\left(\frac{L}{2}\right)\right]}
{\ln\left[\cos\left(\frac{2\pi}{L}\right)\right]} \equiv K_\rho(L).
\label{KrhoL}
\end{eqnarray}
Since the ${\cal O}(L^{-2K_\rho})$ correction in Eq.(\ref{fluctuation}) oscillates on alternate sites, 
the set of $F_L(\frac{L}{2})$ and $F(\frac{L}{2}-2)$ is a more proper choice than that of 
$F_L(\frac{L}{2})$ and $F(\frac{L}{2}-1)$. The values of $K_\rho(L)$ estimated from 
Eq.(\ref{KrhoL}) at $V=0$ as well as near the critical point are shown in Table~\ref{Krho}. 
With increasing $L$, $K_\rho(L)$ seems to approach the exact value 
\begin{eqnarray}
K_\rho=\frac{\pi}{2\arccos[-V/(2t)]}.
\label{Krhoexact}
\end{eqnarray}
We can easily perform the finite-size-scaling analysis with a fitting function $K_\rho(L)=K_\rho+aL^{-\gamma}$ 
($1\lesssim \gamma \le 2$) where $a$ and $\gamma$ are fitting parameters. From Eq.(\ref{KrhoL}), we know 
$a>0$ ($a<0$) for $L/2=$even (odd) and expect a simple relation $\gamma/2 \approx K_\rho$ if the logarithmic 
corrections are small. The fitting analyses are shown in Fig.~\ref{fig5}. As expected, $\gamma/2=1$ 
is obtained at $V/t=0$. Near the critical point, we have $\gamma/2=0.5312$ ($0.4982$) for $V/t=1.8$ ($1.9$) 
and they have a kind of deviation from the exact values of $K_\rho$ due to the existence 
of logarithmic corrections. Nonetheless, the extrapolated values are reasonably close to 
the exact ones (see Table~\ref{Krho}). The remaining error in the thermodynamic limit 
is only about $0.1\%$ ($0.6\%$) for $V/t=1.8$ ($1.9$).

Here, we briefly comment on applying the open boundary conditions (OBC). Thus far the periodic boundary 
conditions are assumed; however, in general the OBC are more preferable in the DMRG study for accurate and 
convenient calculations. In this regard, with applying the OBC the function $S_L(l)$ oscillates like 
Eq.(\ref{fluctuation}) due to the Friedel oscillation and the analysis with Eq.(\ref{cc1}) is unreasonable. 
If one can suppress the Friedel oscillation sufficiently at the center of the system, for example, by using 
the $\sin^2$-deformed OBC technique,~\cite{Gendiar09,Shibata11} the central charge might be calculated 
by taking $S(\frac{L}{2})$ and $S(\frac{L}{2}-2)$ as Eq.(\ref{KrhoL}) even with the OBC.

\section{Conclusion}

A numerical approach to calculate the central charge for 1D quantum system using 
the entanglement entropy is proposed. This approach is applied to the frustrated spin 
chain and the half-filled spinless fermions with nearest-neighbor repulsion to check the validity. 
It is confirmed that the central charge is estimated very accurately for both the 
models even with short system length. We also suggest a new method for determining 
the critical point between TL-liquid and gapped (or ordered) phases by using 
the proposed numerical approach. Furthermore, we demonstrate that the TL parameter 
can be calculated in a like manner as the central charge.

We appreciate Satoshi Ejima for enlightening discussion. We also thank Aroon O'Brien 
and Yohei Fuji for helpful discussions.

\end{document}